\documentclass{birkmult}
\usepackage{amsmath,amssymb,hyperref,bm}
 \newtheorem{thm}{Theorem}[section]
 
 \newtheorem{lem}[thm]{Lemma}
 
 \theoremstyle{definition}
 
 \theoremstyle{remark}

 \numberwithin{equation}{section}

\begin{document}
%
%
%
%
%
%
%
%
%
\title[Dirac equation]
 {Dirac equation as a special case\\ of Cosserat elasticity}
\author[Burnett]{James Burnett}
\address{
Department of Mathematics and Department of Physics \& Astronomy\\
University College London\\
Gower Street\\
London WC1E 6BT\\
UK}
\email{j.burnett@ucl.ac.uk}
\author[Chervova]{Olga Chervova}
\address{
Department of Mathematics\\
University College London\\
Gower Street\\
London WC1E 6BT\\
UK}
\email{olgac@math.ucl.ac.uk}
\author[Vassiliev]{Dmitri Vassiliev}
\address{
Department of Mathematics and Institute of Origins\\
University College London\\
Gower Street\\
London WC1E 6BT\\
UK}
\email{D.Vassiliev@ucl.ac.uk}





\begin{abstract}
We suggest an alternative mathematical model for the electron in
which the dynamical variables are a coframe (field of orthonormal
bases) and a density. The electron mass and external electromagnetic
field are incorporated into our model by means of a Kaluza--Klein
extension. Our Lagrangian density is proportional to axial torsion
squared. The advantage of our approach is that it does not require
the use of spinors, Pauli matrices or covariant differentiation. The
only geometric concepts we use are those of a metric, differential
form, wedge product and exterior derivative. We prove that in the
special case with no dependence on the third spatial coordinate our
model is equivalent to the Dirac equation. The crucial element of
the proof is the observation that our Lagrangian admits a
factorisation.
\end{abstract}

\maketitle
\section{Introduction}

The Dirac equation is a system of four homogeneous linear complex
partial differential equations for four complex unknowns. The
unknowns (components of a bispinor) are functions of time and the
three spatial coordinates. The Dirac equation is the accepted
mathematical model for an electron and its antiparticle, the
positron, in a given external electromagnetic field. One of the main
applications of the Dirac equation is spectral-theoretic: it
determines with high accuracy the energy levels of the hydrogen
atom.

The geometric interpretation of the Dirac equation is rather
complicated. It relies on the use of notions such as
\begin{itemize}
\item
spinor,
\item
Pauli matrices,
\item
covariant derivative (note that formula
(\ref{covariant derivative of spinor field})
for the covariant
derivative of a spinor field is quite tricky).
\end{itemize}
There is also a logical problem with the Dirac equation in that
distinguishing the electron from the positron forces one to resort
to the concept of negative energy. Finally, the electromagnetic
field is incorporated into the Dirac equation by means of a formal
substitution which does not admit a simple geometric interpretation.

The purpose of this paper is to formulate an alternative
mathematical model for the electron and positron, a model which is
geometrically much simpler. The advantage of our approach is that it
does not require the use of spinors, Pauli matrices or covariant
differentiation. The only geometric concepts we use are those of a
\begin{itemize}
\item
metric,
\item
differential form,
\item
wedge product,
\item
exterior derivative.
\end{itemize}
Our model overcomes the logical problem of
distinguishing the electron from the positron: these correspond to
clockwise and anticlockwise rotations of the coframe. And the electromagnetic
field is incorporated into our model by means of a Kaluza--Klein extension
which has a simple geometric interpretation.

The paper has the following structure.
In Section \ref{Notation and conventions} we introduce our notation and
in Section \ref{The Dirac equation} we formulate the Dirac equation.
In Section \ref{Our model} we formulate our mathematical model
and in Section \ref{Choosing a common language} we translate our model
into the language of bispinors. In Section \ref{special case}
we prove Theorem \ref{main theorem} which is the main result of
the paper: this theorem establishes that in the
special case with no dependence on $x^3$
our mathematical model is equivalent to the Dirac equation.
The crucial element of the proof of Theorem \ref{main theorem}
is the observation that our Lagrangian admits a
factorisation; this factorisation is the subject of Lemma \ref{factorisation lemma}.
The concluding discussion is contained in Section~\ref{Discussion}.

\section{Notation and conventions}
\label{Notation and conventions}

Throughout this paper we work on a 4-manifold $M$ equipped with
prescribed Lorentzian metric $g$. All constructions presented in the
paper are local so we do not make a priori assumptions on the
geometric structure of spacetime $\{M,g\}$. The metric $g$ is not
necessarily the Minkowski metric.

Our notation follows \cite{MR2176749,vassilievPRD}.
In particular, in line with the traditions of particle physics,
we use Greek letters to denote tensor (holonomic) indices.

By $\nabla$ we denote the covariant derivative with respect to
the Levi-Civita connection.
It acts on a vector field and a spinor field as
$
\nabla_\alpha v^\beta:=\partial_\alpha v^\beta
+\Gamma^\beta{}_{\alpha\gamma}v^\gamma
$
and
\begin{equation}
\label{covariant derivative of spinor field}
\nabla_\alpha\xi^a:=
\partial_\alpha\xi^a
+\frac14\sigma_\beta{}^{a\dot c}
(\partial_\alpha\sigma^\beta{}_{b\dot c}
+\Gamma^\beta{}_{\alpha\gamma}\sigma^\gamma{}_{b\dot c})\xi^b
\end{equation}
respectively, where
$
\Gamma^\beta{}_{\alpha\gamma}=
\left\{{{\beta}\atop{\alpha\gamma}}\right\}:=
\frac12g^{\beta\delta}
(\partial_\alpha g_{\gamma\delta}
+\partial_\gamma g_{\alpha\delta}
-\partial_\delta g_{\alpha\gamma})
$
are the Christoffel symbols and $\sigma_\beta$ are Pauli matrices.

We identify differential forms with covariant antisymmetric tensors.
Given a pair of real covariant antisymmetric tensors $P$ and $Q$ of
rank $r$ we define their dot product as
$
P\cdot Q:=\frac1{r!}P_{\alpha_1\ldots\alpha_r}Q_{\beta_1\ldots\beta_r}
g^{\alpha_1\beta_1}\ldots g^{\alpha_r\beta_r}
$.
We also define $\|P\|^2:=P\cdot P$.

\section{The Dirac equation}
\label{The Dirac equation}

The following system of linear partial differential equations on $M$
is known as the \emph{Dirac equation}:
\begin{equation}
\label{Dirac equation}
\sigma^{\alpha a\dot b}(i\nabla+A)_\alpha\eta_{\dot b}=m\xi^a,
\qquad
\sigma^\alpha{}_{a\dot b}(i\nabla+A)_\alpha\xi^a=m\eta_{\dot b}\,.
\end{equation}
Here $\xi^a$, $\eta_{\dot b}$ is a bispinor field which plays the role
of dynamical variable (unknown quantity),
$m$ is the electron mass and $A$ is the prescribed electromagnetic covector
potential.
The corresponding Lagrangian density is
\begin{multline}
\label{Dirac Lagrangian}
L_\mathrm{Dir}(\xi,\eta):=
\Bigl[
\frac i2
(\bar\xi^{\dot b}\sigma^\alpha{}_{a\dot b}\nabla_\alpha\xi^a
-
\xi^a\sigma^\alpha{}_{a\dot b}\nabla_\alpha\bar\xi^{\dot b}
+
\bar\eta_a\sigma^{\alpha a\dot b}\nabla_\alpha\eta_{\dot b}
-
\eta_{\dot b}\sigma^{\alpha a\dot b}\nabla_\alpha\bar\eta_a)
\\
+A_\alpha
(\xi^a\sigma^\alpha{}_{a\dot b}\bar\xi^{\dot b}
+\bar\eta_a\sigma^{\alpha a\dot b}\eta_{\dot b})
-m(\xi^a\bar\eta_a+\bar\xi^{\dot b}\eta_{\dot b})
\Bigr]\sqrt{|\det g|}\,.
\end{multline}

\section{Our model}
\label{Our model}

A \emph{coframe} $\vartheta$ is a quartet of real
covector fields
$\vartheta^j$, $j=0,1,2,3$,
satisfying the constraint
\begin{equation}
\label{constraint for coframe}
g=\vartheta^0\otimes\vartheta^0
-\vartheta^1\otimes\vartheta^1
-\vartheta^2\otimes\vartheta^2
-\vartheta^3\otimes\vartheta^3.
\end{equation}
For the sake of clarity we repeat formula~(\ref{constraint for coframe})
giving the tensor indices explicitly:
$g_{\alpha\beta}
=\vartheta^0_\alpha\vartheta^0_\beta
-\vartheta^1_\alpha\vartheta^1_\beta
-\vartheta^2_\alpha\vartheta^2_\beta
-\vartheta^3_\alpha\vartheta^3_\beta$.

Formula (\ref{constraint for coframe}) means that
the coframe is a field of orthonormal bases with
orthonormality understood in the Lorentzian sense.
Of course, at every point of the manifold $M$ the choice of
coframe is not unique: there are 6 real degrees of freedom in
choosing the coframe and any pair of coframes is related by a
Lorentz transformation.

As dynamical variables in our model we choose
a coframe $\vartheta$ and a positive density $\rho$.
These live in the original $(1+3)$-dimensional spacetime
$\{M,g\}$ and are functions of local coordinates
$(x^0,x^1,x^2,x^3)$.

In order to incorporate into our model mass and electromagnetic
field we perform a Kaluza--Klein extension: we add an extra
coordinate $x^4$ and work on the resulting 5-manifold
which we denote by $\mathbf{M}$. We suppose that
\begin{itemize}
\item
the coordinate $x^4$ is fixed,
\item
we allow only changes of coordinates $(x^0,x^1,x^2,x^3)$ which do
not depend on~$x^4$.
\end{itemize}
We will use \textbf{bold} type for extended quantities.

We extend our coframe as
\begin{equation}
\label{Our model equation 1}
{\bm{\vartheta}}{}^0_{\bm{\alpha}}=
\begin{pmatrix}\vartheta^0_\alpha\\0\end{pmatrix},
\qquad
{\bm{\vartheta}}{}^3_{\bm{\alpha}}=
\begin{pmatrix}\vartheta^3_\alpha\\0\end{pmatrix},
\end{equation}
\begin{equation}
\label{Our model equation 2}
({\bm{\vartheta}}{}^1+i{\bm{\vartheta}}{}^2)_{\bm{\alpha}}=
\begin{pmatrix}(\vartheta^1+i\vartheta^2)_\alpha\\0\end{pmatrix}
e^{-2imx^4},
\end{equation}
\begin{equation}
\label{Our model equation 2 point 5}
{\bm{\vartheta}}{}^4_{\bm{\alpha}}=
\begin{pmatrix}0_\alpha\\1\end{pmatrix}
\end{equation}
where the bold tensor index $\bm{\alpha}$ runs through the values 0,
1, 2, 3, 4, whereas its non-bold counterpart $\alpha$ runs through
the values 0, 1, 2, 3. In particular, the $0_\alpha$
in formula (\ref{Our model equation 2 point 5}) stands for a column
of four zeros.

The coordinate $x^4$ parametrises a circle
of radius $\,\frac1{2m}\,$. Condition (\ref{Our model equation 2})
means that the extended coframe $\bm{\vartheta}$
experiences a full turn in the
$(\vartheta^1,\vartheta^2)$-plane as we move along
this circle, coming back to the starting point.

We extend our metric as
\begin{equation}
\label{Our model equation 3}
\mathbf{g}_{{\bm{\alpha}}{\bm{\beta}}}:=
\begin{pmatrix}
g_{\alpha\beta}-\frac1{m^2}A_\alpha A_\beta&\frac1mA_\alpha\\
{}&{}\\
\frac1mA_\beta&-1
\end{pmatrix}.
\end{equation}
Formula (\ref{Our model equation 3}) means that we view
electromagnetism as a perturbation (shear) of the extended metric.
Recall that in classical elasticity ``shear'' stands for ``perturbation of
the metric which does not change the volume''. It is easy to see that
formula (\ref{Our model equation 3}) implies
$\,\det\mathbf{g}=-\det g\,$, so $\,\det\mathbf{g}\,$ does not depend on $A$
and, hence, the electromagnetic field does not change the volume form.

Note that when $A\ne0$ the extended
coframe and the extended metric no longer agree:
\begin{equation}
\label{Our model equation 4}
\mathbf{g}\ne
{\bm{\vartheta}}{}^0\otimes{\bm{\vartheta}}{}^0
-{\bm{\vartheta}}{}^1\otimes{\bm{\vartheta}}{}^1
-{\bm{\vartheta}}{}^2\otimes{\bm{\vartheta}}{}^2
-{\bm{\vartheta}}{}^3\otimes{\bm{\vartheta}}{}^3
-{\bm{\vartheta}}{}^4\otimes{\bm{\vartheta}}{}^4
\end{equation}
(compare with (\ref{constraint for coframe})). The full physical
implications of this discord are not discussed in the current paper.
We need the extended metric only for raising tensor indices
(see formula (\ref{Our model equation 7}) below) and for this purpose
the discord (\ref{Our model equation 4}) is irrelevant.

We define the 3-form
\begin{equation}
\label{Our model equation 5}
\mathbf{T}^\mathrm{ax}:=\frac13
({\bm{\vartheta}}{}^0\wedge d{\bm{\vartheta}}{}^0
-{\bm{\vartheta}}{}^1\wedge d{\bm{\vartheta}}{}^1
-{\bm{\vartheta}}{}^2\wedge d{\bm{\vartheta}}{}^2
-{\bm{\vartheta}}{}^3\wedge d{\bm{\vartheta}}{}^3
-
\underset{=0}
{
\underbrace{
{\bm{\vartheta}}{}^4\!\wedge d{\bm{\vartheta}}{}^4
}
})
\end{equation}
where $\,d\,$ denotes the exterior derivative. This 3-form is called
\emph{axial torsion of the teleparallel connection}. An explanation
of the geometric meaning of the latter phrase as well as a detailed
exposition of the application of torsion in field theory and the
history of the subject can be found in \cite{cartantorsionreview}.
For our purposes the 3-form (\ref{Our model equation 5}) is simply a measure
of deformations generated by rotations of spacetime points.

We choose our Lagrangian density to be
\begin{equation}
\label{Our model equation 6}
L(\vartheta,\rho):=\|\mathbf{T}^\mathrm{ax}\|^2\rho
\end{equation}
where
\begin{equation}
\label{Our model equation 7}
\|\mathbf{T}^\mathrm{ax}\|^2:=\frac1{3!}\,
\mathbf{T}^\mathrm{ax}_{{\bm{\alpha}}{\bm{\beta}}{\bm{\gamma}}}\,
\mathbf{T}^\mathrm{ax}_{{\bm{\kappa}}{\bm{\lambda}}{\bm{\mu}}}\,
\mathbf{g}^{{\bm{\alpha}}{\bm{\kappa}}}\,
\mathbf{g}^{{\bm{\beta}}{\bm{\lambda}}}\,
\mathbf{g}^{{\bm{\gamma}}{\bm{\mu}}}\,.
\end{equation}

Formula (\ref{Our model equation 2}) implies
\begin{equation}
\label{Our model equation 8}
{\bm{\vartheta}}{}^1\wedge d{\bm{\vartheta}}{}^1+
{\bm{\vartheta}}{}^2\wedge d{\bm{\vartheta}}{}^2=
\vartheta^1\wedge d\vartheta^1+\vartheta^2\wedge d\vartheta^2
-4m\vartheta^1\wedge\vartheta^2\wedge{\bm{\vartheta}}{}^4
\end{equation}
so our Lagrangian density $L(\vartheta,\rho)$ does not depend on
$x^4$ and can be viewed as a Lagrangian density in the original
spacetime of dimension $1+3$. This means, essentially, that we have
performed a separation of variables in a nonlinear setting.

Our action (variational functional) is
$\int L(\vartheta,\rho)\,dx^0dx^1dx^2dx^3$.
Our field equations (Euler--Lagrange equations) are obtained by
varying this action with respect to the coframe $\vartheta$ and
density $\rho$. Varying with respect to the density $\rho$ is easy:
this gives the field equation $\|\mathbf{T}^\mathrm{ax}\|^2=0$ which is
equivalent to $L(\vartheta,\rho)=0$. Varying with respect to the
coframe $\vartheta$ is more difficult because we have to maintain
the metric constraint (\ref{constraint for coframe}); recall that
the metric is assumed to be prescribed (fixed).

We do not write down the field equations for the Lagrangian density
$L(\vartheta,\rho)$ explicitly. We note only that they are highly
nonlinear and do not appear to bear any resemblance to the linear Dirac
equation (\ref{Dirac equation}).

\section{Choosing a common language}
\label{Choosing a common language}

In order to compare the two models described in Sections \ref{The
Dirac equation} and \ref{Our model} we need to choose a common
mathematical language. We choose the language of bispinors.
Namely, we express the coframe and density via a bispinor field
according to formulae
\begin{equation}
\label{common language equation 1}
s=\xi^a\bar\eta_a,
\end{equation}
\begin{equation}
\label{common language equation 2}
\rho=|s|\,\sqrt{\det|g_{\alpha\beta}|}\,,
\end{equation}
\begin{equation}
\label{common language equation 3}
(\vartheta^0+\vartheta^3)_\alpha
=|s|^{-1}\xi^a\sigma_{\alpha a\dot b}\bar\xi^{\dot b},
\end{equation}
\begin{equation}
\label{common language equation 4}
(\vartheta^0-\vartheta^3)_\alpha
=|s|^{-1}\bar\eta^a\sigma_{\alpha a\dot b}\eta^{\dot b},
\end{equation}
\begin{equation}
\label{common language equation 5}
(\vartheta^1+i\vartheta^2)_\alpha
=-|s|^{-1}\xi^a\sigma_{\alpha a\dot b}\eta^{\dot b}
\end{equation}
where
\begin{equation}
\label{common language equation 6}
\eta^{\dot a}=\epsilon^{\dot a\dot b}\eta_{\dot b},
\qquad
\epsilon_{ab}=\epsilon_{\dot a\dot b}=
\epsilon^{ab}=\epsilon^{\dot a\dot b}=
\begin{pmatrix}
0&1\\
-1&0
\end{pmatrix}.
\end{equation}
Note that throughout this paper we assume that the density $\rho$
does not vanish.

Observe now that the right-hand sides of formulae
(\ref{common language equation 2})--(\ref{common language equation 5})
are invariant under the change of bispinor field
$\xi^a\mapsto\xi^ae^{i\varphi}$,
$\eta_{\dot b}\mapsto\eta_{\dot b}e^{-i\varphi}$
where $\varphi:M\to\mathbb{R}$ is an arbitrary scalar function.
In other words, formulae
(\ref{common language equation 2})--(\ref{common language equation 5})
do not feel the argument of the complex scalar $s$.
Hence, when translating our model into the language of bispinors it is
natural to impose the constraint
\begin{equation}
\label{common language equation 7}
\operatorname{Im}s=0,\qquad s>0.
\end{equation}
This constraint
reflects the fact that
our model has one real dynamical degree of freedom less than the Dirac model
(seven real degrees of freedom instead of eight).

\section{Special case with no dependence on $x^3$}
\label{special case}

In addition to our usual assumptions
(see beginning of Section~\ref{Our model})
we suppose that
\begin{itemize}
\item
the coordinate $x^3$ is fixed,
\item
we allow only changes of coordinates $(x^0,x^1,x^2)$ which do
not depend on~$x^3$,
\item
the metric does not depend on $x^3$ and has block structure
\begin{equation}
\label{special case equation 1}
g_{\alpha\beta}=
\begin{pmatrix}
g_{00}&g_{01}&g_{02}&0\\
g_{10}&g_{11}&g_{12}&0\\
g_{20}&g_{21}&g_{22}&0\\
0&0&0&-1\\
\end{pmatrix},
\end{equation}
\item
the electromagnetic covector potential does not depend on $x^3$ and has
$A_3=0$.
\
\end{itemize}

We work with coframes such that
\begin{equation}
\label{special case equation 2}
\vartheta^3_\alpha=
\begin{pmatrix}
0\\
0\\
0\\
1
\end{pmatrix}.
\end{equation}

We use Pauli matrices which do not depend on $x^3$ and take
\begin{equation}
\label{special case equation 3}
\sigma_{3a\dot b}=
\begin{pmatrix}
1&0\\
0&-1\end{pmatrix}.
\end{equation}

We take
\begin{equation}
\label{special case equation 4}
\eta_{\dot b}=\xi^a\sigma_{3a\dot b}\,.
\end{equation}
Then the scalar defined by formula
(\ref{common language equation 1}) takes the form $s=|\xi^1|^2-|\xi^2|^2$.
This scalar is automatically real and condition
(\ref{common language equation 7}) becomes
\begin{equation}
\label{special case equation 5}
|\xi^1|^2-|\xi^2|^2>0.
\end{equation}
It is easy to see that formulae
(\ref{special case equation 1}),
(\ref{special case equation 3})--(\ref{special case equation 5})
imply (\ref{special case equation 2}).

Formula (\ref{special case equation 4}) means that our bispinor
$\xi^a$, $\eta_{\dot b}$
is determined by the spinor $\xi^a$.
Thus, the spinor $\xi^a$ becomes the (only) dynamical variable.
We assume that this spinor does not depend on $x^3$.

Observe that in the special case considered in this section both the
Dirac model and our model have the same number of real dynamical
degrees of freedom, namely, four. This is because under the assumption
(\ref{special case equation 2}) the coframe $\vartheta$ and
density $\rho$ are equivalent to a spinor field $\xi^a$ modulo sign
($-\xi^a$ gives the same $\vartheta$ and $\rho$).

Throughout this section summation is carried out either over indices 0, 1, 2
or over indices 0, 1, 2, 4. In the latter case we use
\textbf{bold} type.

Put
\begin{multline}
\label{Dirac Lagrangian plus minus}
L_\mathrm{Dir}^\pm(\xi):=
\Bigl[
\frac i2
(\bar\xi^{\dot b}\sigma^\alpha{}_{a\dot b}\nabla_\alpha\xi^a
-
\xi^a\sigma^\alpha{}_{a\dot b}\nabla_\alpha\bar\xi^{\dot b})
\\
+A_\alpha
\xi^a\sigma^\alpha{}_{a\dot b}\bar\xi^{\dot b}
\mp m\xi^a\sigma_{3a\dot b}\bar\xi^{\dot b}
\Bigr]\sqrt{|\det g|}\,.
\end{multline}
The Lagrangian densities $L_\mathrm{Dir}^\pm(\xi)$ are
formally related to the original Lagrangian density
(\ref{Dirac Lagrangian}) as follows: if we set
$\eta_{\dot b}=\pm\xi^a\sigma_{3a\dot b}$ we get
$L_\mathrm{Dir}(\xi,\eta)=2L_\mathrm{Dir}^\pm(\xi)$.
We say ``formally related'' because in this section we assume
that formula $\eta_{\dot b}=\pm\xi^a\sigma_{3a\dot b}$ holds with upper sign,
see (\ref{special case equation 4}). The
$L_\mathrm{Dir}^+(\xi)$ and $L_\mathrm{Dir}^-(\xi)$
are, of course, the usual Dirac Lagrangian densities for an electron
with spin up and spin down.

\begin{lem}
\label{factorisation lemma}
In the special case with no dependence on $x^3$
our Lagrangian density~(\ref{Our model equation 6}) factorises as
\begin{equation}
\label{factorisation}
L(\vartheta,\rho)=-\frac{32m}9
\frac{L_\mathrm{Dir}^+(\xi)L_\mathrm{Dir}^-(\xi)}
{L_\mathrm{Dir}^+(\xi)-L_\mathrm{Dir}^-(\xi)}\,.
\end{equation}
\end{lem}

Let us emphasise once again that throughout this paper we assume
that the density $\rho$ does not vanish. In the special case
with no dependence on $x^3$
this assumption can be
equivalently rewritten as
\begin{equation}
\label{density is nonzero}
L_\mathrm{Dir}^+(\xi)\ne L_\mathrm{Dir}^-(\xi)
\end{equation}
so the denominator in (\ref{factorisation}) is nonzero.

\begin{proof}
\textbf{Step 1.} Let us show that it is sufficient to prove formula
(\ref{factorisation}) under the assumption $dA=0$, i.e. under the
assumption that the electromagnetic covector potential $A$ is pure
gauge. Recall that $dA$ stands for the exterior derivative of $A$.

Suppose that we have already proved formula (\ref{factorisation}) under
the assumption $dA=0$ and are now looking at the case of general
$A$. Let us fix an arbitrary point $P$ on our 4-manifold $M$
and prove formula (\ref{factorisation}) at this point. To
do this, we perturb the electromagnetic covector potential $A$ in
such a way that
\begin{itemize}
\item
$A$ retains its value at the point $P$ and
\item
$A$ satisfies the condition $dA=0$ in a neighbourhood of $P$.
\end{itemize}
This can be achieved by, say, choosing some local coordinates on $M$
and setting the components of $A$ to be constant in this coordinate
system. Now, this perturbation of the covector potential $A$ does
not change the LHS or the RHS of (\ref{factorisation}) at the point
$P$ because neither of them depends on derivatives of $A$.
Hence, the case of general $A$ has been reduced to the case $dA=0$.

\textbf{Step 2.}
Let us show that it is sufficient to prove formula
(\ref{factorisation}) under the assumption $A=0$.

Suppose that we have already proved formula (\ref{factorisation}) under
the assumption $A=0$ and are now looking at the case
$dA=0$. Let us modify the definition of the extended coframe by replacing
(\ref{Our model equation 2 point 5}) with
\begin{equation}
\label{proof of lemma equation 1}
{\bm{\vartheta}}{}^4_{\bm{\alpha}}=
\begin{pmatrix}-\frac1mA_\alpha\\1\end{pmatrix}.
\end{equation}
In view of the condition $dA=0$ this modification of the extended
coframe does not change axial torsion (\ref{Our model equation 5})
but the extended coframe
(\ref{Our model equation 1}),
(\ref{Our model equation 2}),
(\ref{proof of lemma equation 1})
now agrees with the extended metric
(\ref{Our model equation 3}): we have
\begin{equation}
\label{proof of lemma equation 2}
\mathbf{g}=
{\bm{\vartheta}}{}^0\otimes{\bm{\vartheta}}{}^0
-{\bm{\vartheta}}{}^1\otimes{\bm{\vartheta}}{}^1
-{\bm{\vartheta}}{}^2\otimes{\bm{\vartheta}}{}^2
-{\bm{\vartheta}}{}^3\otimes{\bm{\vartheta}}{}^3
-{\bm{\vartheta}}{}^4\otimes{\bm{\vartheta}}{}^4
\end{equation}
as opposed to (\ref{Our model equation 4}). Let us now perform a change of
coordinates
\begin{equation}
\label{proof of lemma equation 3}
\tilde x^\alpha=x^\alpha,\quad\alpha=0,1,2,3,
\qquad
\tilde x^4=x^4-\frac1m\int A\cdot dx.
\end{equation}
Note that the integral $\int A\cdot dx$ is (locally) well-defined because
of the assumption $dA=0$.
The change of coordinates
(\ref{proof of lemma equation 3})
is against the rules we stated in
the beginning of Section \ref{Our model} when describing our model
(we changed
the original Kaluza coordinate $x^4$ to a new coordinate $\tilde x^4$)
but we are doing this only for the purpose of proving the lemma.
In the new coordinate system $\tilde x$ the extended coframe
(\ref{Our model equation 1}),
(\ref{Our model equation 2}),
(\ref{proof of lemma equation 1})
takes its original form
(\ref{Our model equation 1})--(\ref{Our model equation 2 point 5}),
the extended metric takes the form
$
\mathbf{g}_{{\bm{\alpha}}{\bm{\beta}}}=
\begin{pmatrix}
g_{\alpha\beta}&0\\
{}&{}\\
0&-1
\end{pmatrix}
$
(compare with (\ref{Our model equation 3}))
and the electromagnetic covector potential $A$ is not affected
(i.e. it has the same components in both coordinate systems).
Observe now that in (\ref{Our model equation 2}) we
have retained the scalar factor
$e^{-2imx^4}$ written in terms of the original Kaluza coordinate $x^4$.
Expressing $x^4$ in terms of $\tilde x^4$ in accordance with
formula (\ref{proof of lemma equation 3}) we get
\begin{equation}
\label{proof of lemma equation 5}
({\bm{\vartheta}}{}^1+i{\bm{\vartheta}}{}^2)_{\bm{\alpha}}=
\begin{pmatrix}(\vartheta^1+i\vartheta^2)_\alpha\\0\end{pmatrix}
e^{-2im\tilde x^4-2i\int A\cdot dx}.
\end{equation}

Let us now introduce a
new coframe $\hat\vartheta$ in $(1+3)$-dimensional spacetime
$\{M,g\}$
related to the original coframe $\vartheta$ as
\begin{equation}
\label{proof of lemma equation 6}
\hat\vartheta^0=\vartheta^0,\qquad
\hat\vartheta^3=\vartheta^3,\qquad
\hat\vartheta^1+i\hat\vartheta^2=(\vartheta^1+i\vartheta^2)
e^{-2i\int A\cdot dx}.
\end{equation}
Then formulae (\ref{proof of lemma equation 5}),
(\ref{proof of lemma equation 6}) imply
\begin{equation}
\label{proof of lemma equation 7}
L(\hat\vartheta,\rho;0)=L(\vartheta,\rho;A).
\end{equation}
Here $L(\,\cdot\,,\,\cdot\,;\,\cdot\,)$ is our Lagrangian density
$L(\,\cdot\,,\,\cdot\,)$
defined by formulae
(\ref{Our model equation 1})--(\ref{Our model equation 3}),
(\ref{Our model equation 5})--(\ref{Our model equation 7})
but with an extra entry after the semicolon for
the electromagnetic covector potential.
Formula (\ref{proof of lemma equation 7}) means that in our model
the introduction of an electromagnetic covector potential $A$
satisfying the condition $dA=0$ is equivalent to a change of coframe
(\ref{proof of lemma equation 6}).

Formulae
(\ref{common language equation 1})--(\ref{common language equation 6}),
(\ref{special case equation 4})
imply that the change of coframe (\ref{proof of lemma equation 6})
leads to a change of spinor field
$
\hat\xi^a=\xi^ae^{-i\int A\cdot dx}
$.
Substituting the latter into
(\ref{Dirac Lagrangian plus minus}) we get
\begin{equation}
\label{proof of lemma equation 9}
L_\mathrm{Dir}^\pm(\hat\xi;0)=L_\mathrm{Dir}^\pm(\xi;A).
\end{equation}
Here $L_\mathrm{Dir}^\pm(\,\cdot\,;\,\cdot\,)$ is the Dirac Lagrangian density
$L_\mathrm{Dir}^\pm(\,\cdot\,)$ defined by formula
(\ref{Dirac Lagrangian plus minus})
but with an extra entry after the semicolon for
the electromagnetic covector potential.

In the beginning of this part of the proof we assumed that we have
already proved formula (\ref{factorisation}) under the assumption
$A=0$ so we have
\begin{equation}
\label{proof of lemma equation 10}
L(\hat\vartheta,\rho;0)=-\frac{32m}9
\frac{L_\mathrm{Dir}^+(\hat\xi;0)L_\mathrm{Dir}^-(\hat\xi;0)}
{L_\mathrm{Dir}^+(\hat\xi;0)-L_\mathrm{Dir}^-(\hat\xi;0)}\,.
\end{equation}
It remains to note that formulae
(\ref{proof of lemma equation 7})--(\ref{proof of lemma equation 10})
imply (\ref{factorisation}).
Hence, the case $dA=0$ has been reduced to the case $A=0$.

\textbf{Step 3.}
In the remainder of the proof we assume that $A=0$.

The proof of formula (\ref{factorisation}) is performed by direct
substitution: it is just a matter of expressing the coframe and
density via the spinor using formulae
(\ref{common language equation 1})--(\ref{common language equation 6}),
(\ref{special case equation 4})
and substituting these expressions into the LHS of
(\ref{factorisation}). However, even with $A=0$ this is a massive
calculation. In order to overcome these technical difficulties we
perform below a trick which makes the calculations much easier.
This trick is a known one and was, for example, extensively used by
A.~Dimakis and F.~M\"uller-Hoissen
\cite{muellerhoissen1,muellerhoissen2,muellerhoissen3}.

Observe that when working with spinors we have the
freedom in our choice of Pauli matrices: at every point of our
$(1+3)$-dimensional spacetime $\{M,g\}$
we can apply a proper Lorentz transformation to a given set of Pauli
matrices to get a new set of Pauli matrices, with spinor fields
transforming accordingly. It is sufficient to prove formula
(\ref{factorisation}) for one particular choice of Pauli matrices,
hence it is natural to choose Pauli matrices in a way that makes
calculations as simple as possible. We choose Pauli matrices
\begin{equation}
\label{special formula for Pauli matrices}
\sigma_{\alpha a\dot b}=\vartheta^j_\alpha\,s_{ja\dot b}
=\vartheta^0_\alpha\,s_{0a\dot b}
+\vartheta^1_\alpha\,s_{1a\dot b}
+\vartheta^2_\alpha\,s_{2a\dot b}
+\vartheta^3_\alpha\,s_{3a\dot b}
\end{equation}
where
\begin{equation}
\label{Pauli matrices s}
s_{ja\dot b}=
\begin{pmatrix}
s_{0a\dot b}\\
s_{1a\dot b}\\
s_{2a\dot b}\\
s_{3a\dot b}
\end{pmatrix}
:=
\begin{pmatrix}
\begin{pmatrix}
1&0\\
0&1
\end{pmatrix}
\\
\begin{pmatrix}
0&1\\
1&0
\end{pmatrix}
\\
\begin{pmatrix}
0&i\\
-i&0
\end{pmatrix}
\\
\begin{pmatrix}
1&0\\
0&-1\end{pmatrix}
\end{pmatrix}.
\end{equation}
Here $\vartheta$ is the coframe that appears in the LHS of formula
(\ref{factorisation}). Let us stress that in the statement of the
lemma Pauli matrices are not assumed to be related in any way to the
coframe $\vartheta$. We are just choosing the particular Pauli
matrices (\ref{special formula for Pauli matrices}),
(\ref{Pauli matrices s}) to simplify calculations in our proof.

Examination of formulae
(\ref{common language equation 1})--(\ref{common language equation 6}),
(\ref{special case equation 4}), (\ref{special case equation 5}),
(\ref{special formula for Pauli matrices}),
(\ref{Pauli matrices s})
shows that with our special choice of Pauli matrices
we have $\xi^2=0$ whereas $\xi^1$ is nonzero and real.
We are about to write down the
Dirac Lagrangian density
(\ref{Dirac Lagrangian plus minus})
which is quadratic in $\xi$ so the sign of
$\xi$ does not matter.
So let
$\xi^a=
\begin{pmatrix}
e^h\\0
\end{pmatrix}$
where $h:M\to\mathbb{R}$ is a scalar function.
We get
\begin{multline*}
\frac i2
\bar\xi^{\dot d}\sigma^\alpha{}_{a\dot d}\nabla_\alpha\xi^a
=\frac i2
\bar\xi^{\dot d}(\sigma^\alpha{}_{a\dot d}\partial_\alpha h)\xi^a
+\frac i8\bar\xi^{\dot d}\sigma^\alpha{}_{a\dot d}
\sigma_\beta{}^{a\dot c}
(\partial_\alpha\sigma^\beta{}_{b\dot c}
+\Gamma^\beta{}_{\alpha\gamma}\sigma^\gamma{}_{b\dot c})\xi^b
\\
=\frac{ie^{2h}}8\sigma^\alpha{}_{a\dot 1}
\sigma_\beta{}^{a\dot c}
(\partial_\alpha\sigma^\beta{}_{1\dot c}
+\Gamma^\beta{}_{\alpha\gamma}\sigma^\gamma{}_{1\dot c})+\ldots
=\frac{ie^{2h}}8\sigma^\alpha{}_{a\dot 1}
\sigma_\beta{}^{a\dot c}
\nabla_\alpha\sigma^\beta{}_{1\dot c}+\ldots
\\
=\frac{ie^{2h}}8
\bigl[\sigma^\alpha{}_{1\dot 1}
\sigma_\beta{}^{1\dot 1}
\nabla_\alpha\sigma^\beta{}_{1\dot 1}
+\sigma^\alpha{}_{1\dot 1}
\sigma_\beta{}^{1\dot 2}
\nabla_\alpha\sigma^\beta{}_{1\dot 2}
+\sigma^\alpha{}_{2\dot 1}
\sigma_\beta{}^{2\dot 1}
\nabla_\alpha\sigma^\beta{}_{1\dot 1}
+\sigma^\alpha{}_{2\dot 1}
\sigma_\beta{}^{2\dot 2}
\nabla_\alpha\sigma^\beta{}_{1\dot 2}\bigr]+\ldots
\\
\!\!\!\!\!\!\!\!\!\!\!\!\!\!\!\!\!\!\!\!\!\!\!\!\!\!\!\!\!\!\!\!\!\!\!\!
\!\!\!\!\!\!\!\!\!\!\!\!\!\!\!\!\!\!\!\!\!\!\!\!\!\!\!\!\!\!\!\!\!\!\!\!
\!\!\!\!\!\!\!\!\!\!\!\!\!\!\!\!\!\!\!\!\!\!\!\!\!\!\!\!
=\frac{ie^{2h}}8
\bigl[\vartheta^{0\alpha}
\sigma_\beta{}^{1\dot 1}
\nabla_\alpha\sigma^\beta{}_{1\dot 1}
+\vartheta^{0\alpha}
\sigma_\beta{}^{1\dot 2}
\nabla_\alpha\sigma^\beta{}_{1\dot 2}
\\
\qquad\qquad\qquad\qquad\qquad\qquad\qquad
+(\vartheta^1-i\vartheta^2)^\alpha
\sigma_\beta{}^{2\dot 1}
\nabla_\alpha\sigma^\beta{}_{1\dot 1}
+(\vartheta^1-i\vartheta^2)^\alpha
\sigma_\beta{}^{2\dot 2}
\nabla_\alpha\sigma^\beta{}_{1\dot 2}\bigr]+\ldots
\\
\!\!\!\!\!\!\!\!\!\!\!\!\!\!\!\!\!\!\!\!\!\!\!\!\!\!\!\!\!\!\!\!\!\!\!\!
\!\!\!\!\!\!\!\!\!\!\!\!\!\!\!\!\!\!\!\!\!\!\!\!\!\!\!\!\!\!\!\!\!\!\!\!
\!\!\!\!\!\!\!\!\!\!\!\!\!
=\frac{ie^{2h}}8
\bigl[\vartheta^{0\alpha}
\sigma_\beta{}^{1\dot 1}
\nabla_\alpha\vartheta^{0\beta}
+\vartheta^{0\alpha}
\sigma_\beta{}^{1\dot 2}
\nabla_\alpha(\vartheta^1+i\vartheta^2)^\beta
\\
\qquad\qquad\qquad\qquad\qquad\qquad
+(\vartheta^1-i\vartheta^2)^\alpha
\sigma_\beta{}^{2\dot 1}
\nabla_\alpha\vartheta^{0\beta}
+(\vartheta^1-i\vartheta^2)^\alpha
\sigma_\beta{}^{2\dot 2}
\nabla_\alpha(\vartheta^1+i\vartheta^2)^\beta\bigr]+\ldots
\\
\!\!\!\!\!\!\!\!\!\!\!\!\!\!\!\!\!\!\!\!\!\!\!\!\!\!\!\!\!\!\!\!\!\!\!\!
\!\!\!\!\!\!\!\!\!\!\!\!\!\!\!\!\!\!\!\!\!\!\!\!\!
\!\!\!\!\!\!\!\!\!\!\!
=\frac{ie^{2h}}8
\bigl[\vartheta^{0\alpha}
\vartheta^0_\beta
\nabla_\alpha\vartheta^{0\beta}
-\vartheta^{0\alpha}
(\vartheta^1-i\vartheta^2)_\beta
\nabla_\alpha(\vartheta^1+i\vartheta^2)^\beta
\\
\qquad\qquad\qquad\qquad\qquad
-(\vartheta^1-i\vartheta^2)^\alpha
(\vartheta^1+i\vartheta^2)_\beta
\nabla_\alpha\vartheta^{0\beta}
+(\vartheta^1-i\vartheta^2)^\alpha
\vartheta^0_\beta
\nabla_\alpha(\vartheta^1+i\vartheta^2)^\beta\bigr]+\ldots
\\
\!\!\!\!\!\!\!\!\!\!\!\!\!\!\!\!\!\!\!\!\!\!\!\!\!\!\!\!\!\!\!\!\!\!\!\!
\!\!\!\!\!\!\!\!\!\!\!\!\!\!\!\!\!\!\!\!\!\!\!\!\!
\!\!\!
=\frac{ie^{2h}}8
\bigl[
-i\vartheta^{0\alpha}\vartheta^1_\beta\nabla_\alpha\vartheta^{2\beta}
+i\vartheta^{0\alpha}\vartheta^2_\beta\nabla_\alpha\vartheta^{1\beta}
-i\vartheta^{1\alpha}\vartheta^2_\beta\nabla_\alpha\vartheta^{0\beta}
\\
\qquad\qquad\qquad\qquad\qquad\qquad\qquad\quad
+i\vartheta^{2\alpha}\vartheta^1_\beta\nabla_\alpha\vartheta^{0\beta}
+i\vartheta^{1\alpha}\vartheta^0_\beta\nabla_\alpha\vartheta^{2\beta}
-i\vartheta^{2\alpha}\vartheta^0_\beta\nabla_\alpha\vartheta^{1\beta}\bigr]
+\ldots
\\
\!\!\!\!\!\!\!\!\!\!\!\!\!\!\!\!\!\!\!\!\!\!\!\!\!\!\!\!\!\!\!\!\!\!\!\!
\!\!\!\!\!\!\!\!\!\!\!\!\!\!\!\!\!\!\!\!\!\!\!\!\!
\!\!\!\!\!\!\!\!\!\!\!\!\!\!
=\frac{e^{2h}}8
\bigl[\vartheta^{0\alpha}\vartheta^1_\beta\nabla_\alpha\vartheta^{2\beta}
-\vartheta^{0\alpha}\vartheta^2_\beta\nabla_\alpha\vartheta^{1\beta}
+\vartheta^{1\alpha}\vartheta^2_\beta\nabla_\alpha\vartheta^{0\beta}
\\
\qquad\qquad\qquad\qquad\qquad\qquad\qquad\qquad
-\vartheta^{2\alpha}\vartheta^1_\beta\nabla_\alpha\vartheta^{0\beta}
-\vartheta^{1\alpha}\vartheta^0_\beta\nabla_\alpha\vartheta^{2\beta}
+\vartheta^{2\alpha}\vartheta^0_\beta\nabla_\alpha\vartheta^{1\beta}\bigr]
+\ldots
\\
=\frac s8
\bigl[(\vartheta^0\wedge\vartheta^1)\cdot d\vartheta^2
+(\vartheta^1\wedge\vartheta^2)\cdot d\vartheta^0
+(\vartheta^2\wedge\vartheta^0)\cdot d\vartheta^1\bigr]
+\ldots
\end{multline*}
where the dots denote purely imaginary terms.
Hence,
\[
\frac i2
(\bar\xi^{\dot b}\sigma^\alpha{}_{a\dot b}\nabla_\alpha\xi^a
-
\xi^a\sigma^\alpha{}_{a\dot b}\nabla_\alpha\bar\xi^{\dot b})
=\frac s4
[(\vartheta^0\wedge\vartheta^1)\cdot d\vartheta^2
+(\vartheta^1\wedge\vartheta^2)\cdot d\vartheta^0
+(\vartheta^2\wedge\vartheta^0)\cdot d\vartheta^1].
\]
Formula (\ref{Dirac Lagrangian plus minus}) with $A=0$
can now be rewritten as
\begin{equation}
\label{nice formula for Dirac Lagrangian in dimension 1+2}
L_\mathrm{Dir}^\pm(\xi)=
\left[
\frac14
\bigl[(\vartheta^0\wedge\vartheta^1)\cdot d\vartheta^2
+(\vartheta^1\wedge\vartheta^2)\cdot d\vartheta^0
+(\vartheta^2\wedge\vartheta^0)\cdot d\vartheta^1\bigr]
\mp m
\right]\rho\,.
\end{equation}

Put
\begin{equation}
\label{proof of lemma final stage equation 1}
T^\mathrm{ax}:=\frac13
(\vartheta^0\wedge d\vartheta^0
-\vartheta^1\wedge d\vartheta^1
-\vartheta^2\wedge d\vartheta^2
-
\underset{=0}
{
\underbrace{
\vartheta^3\wedge d\vartheta^3
}
})
\end{equation}
(compare with (\ref{Our model equation 5})).
The last term in (\ref{proof of lemma final stage equation 1})
vanishes in view of (\ref{special case equation 2}).
The coordinate $x^3$ is redundant so
$T^\mathrm{ax}$ can be viewed as a 3-form in
$(1+2)$-dimensional Lorentzian space
with local coordinates $(x^0,x^1,x^2)$.
Hence, we can define the scalar
\begin{equation}
\label{proof of lemma final stage equation 2}
*T^\mathrm{ax}:=
\frac1{3!}\,\sqrt{|\det g|}\,
(T^\mathrm{ax})^{\alpha\beta\gamma}\varepsilon_{\alpha\beta\gamma}
\end{equation}
which is the Hodge dual of $T^\mathrm{ax}$.
But $\sqrt{|\det g|}\,\varepsilon_{\alpha\beta\gamma}
=(\vartheta^0\wedge\vartheta^1\wedge\vartheta^2)_{\alpha\beta\gamma}$
so formula
(\ref{proof of lemma final stage equation 2}) can be rewritten as
\begin{multline*}
*T^\mathrm{ax}
=T^\mathrm{ax}
\cdot(\vartheta^0\wedge\vartheta^1\wedge\vartheta^2)
=\frac13
(\vartheta^0\wedge d\vartheta^0
-\vartheta^1\wedge d\vartheta^1
-\vartheta^2\wedge d\vartheta^2)
\cdot(\vartheta^0\wedge\vartheta^1\wedge\vartheta^2)
\\
=\frac13
\bigl[(\vartheta^0\wedge\vartheta^1)\cdot d\vartheta^2
+(\vartheta^1\wedge\vartheta^2)\cdot d\vartheta^0
+(\vartheta^2\wedge\vartheta^0)\cdot d\vartheta^1\bigr].
\end{multline*}
Substituting the latter into
(\ref{nice formula for Dirac Lagrangian in dimension 1+2})
we arrive at the compact formula
\begin{equation}
\label{proof of lemma final stage equation 4}
L_\mathrm{Dir}^\pm(\xi)=
\left[
\frac34*T^\mathrm{ax}
\mp m
\right]\rho\,.
\end{equation}
Substituting
(\ref{proof of lemma final stage equation 4}) into the RHS of
(\ref{factorisation}) we get
\begin{equation*}
-\frac{32m}9
\frac{L_\mathrm{Dir}^+(\xi)L_\mathrm{Dir}^-(\xi)}
{L_\mathrm{Dir}^+(\xi)-L_\mathrm{Dir}^-(\xi)}
=\left[(*T^\mathrm{ax})^2-\frac{16}9m^2\right]\rho
\,.
\end{equation*}
As our Lagrangian $L(\vartheta,\rho)$ is defined by formula
(\ref{Our model equation 6}), the proof of the lemma has been reduced to
proving
\begin{equation}
\label{proof of lemma final stage equation 6}
\|\mathbf{T}^\mathrm{ax}\|^2
=(*T^\mathrm{ax})^2-\frac{16}9m^2
\end{equation}
with $A=0$ (recall that $A$ initially appeared in the extended
metric (\ref{Our model equation 3})).

In view of
(\ref{Our model equation 1}),
(\ref{special case equation 2}) formula
(\ref{Our model equation 5}) becomes
\begin{equation}
\label{proof of lemma final stage equation 7}
\mathbf{T}^\mathrm{ax}=\frac13
({\bm{\vartheta}}{}^0\wedge d{\bm{\vartheta}}{}^0
-{\bm{\vartheta}}{}^1\wedge d{\bm{\vartheta}}{}^1
-{\bm{\vartheta}}{}^2\wedge d{\bm{\vartheta}}{}^2).
\end{equation}
The difference between formulae (\ref{proof of lemma final stage equation 1})
and (\ref{proof of lemma final stage equation 7}) is that
the RHS of (\ref{proof of lemma final stage equation 1})
is expressed via the coframe $\vartheta$ in the original
spacetime
whereas the RHS of (\ref{proof of lemma final stage equation 7})
is expressed via the coframe $\bm{\vartheta}$ in the extended
spacetime, see
(\ref{Our model equation 1})--(\ref{Our model equation 2 point 5}).
In view of (\ref{Our model equation 8}),
(\ref{proof of lemma final stage equation 1})
formula (\ref{proof of lemma final stage equation 7}) can be rewritten as
\begin{equation}
\label{proof of lemma final stage equation 8}
\mathbf{T}^\mathrm{ax}
=T^\mathrm{ax}
+\frac{4m}3\vartheta^1\wedge\vartheta^2\wedge{\bm{\vartheta}}{}^4.
\end{equation}
The coordinate $x^3$ is redundant so
$\mathbf{T}^\mathrm{ax}$ can be viewed as a 3-form in
$(1+3)$-dimensional Lorentzian space
with local coordinates $(x^0,x^1,x^2,x^4)$.
Hence, we can define the covector
\begin{equation}
\label{proof of lemma final stage equation 9}
(*\mathbf{T}^\mathrm{ax})_{\bm{\delta}}:=
\frac1{3!}\,\sqrt{|\det g|}\,
(\mathbf{T}^\mathrm{ax})^{{\bm{\alpha}}{\bm{\beta}}{\bm{\gamma}}}
\varepsilon_{{\bm{\alpha}}{\bm{\beta}}{\bm{\gamma}}{\bm{\delta}}}\,,
\qquad\bm{\delta}=0,1,2,4,
\end{equation}
which is the Hodge dual of $\mathbf{T}^\mathrm{ax}$.
It is easy to see that we have
\begin{equation}
\label{proof of lemma final stage equation 10}
\|\mathbf{T}^\mathrm{ax}\|^2=-\|*\mathbf{T}^\mathrm{ax}\|^2.
\end{equation}
Note that in the LHS of (\ref{proof of lemma final stage equation 10})
we square a 3-form in $(1+4)$-dimensional Lorentzian space
whereas in the RHS of (\ref{proof of lemma final stage equation 10})
we square a 1-form in $(1+3)$-dimensional Lorentzian space,
so we took great care in getting the sign right.
Substituting
(\ref{proof of lemma final stage equation 8})
into
(\ref{proof of lemma final stage equation 9})
we get
\begin{equation}
\label{proof of lemma final stage equation 11}
(*\mathbf{T}^\mathrm{ax})_{\bm{\delta}}
=\begin{pmatrix}
\frac{4m}3\vartheta^0_\delta\\
*{T}^\mathrm{ax}
\end{pmatrix}
\end{equation}
where $*{T}^\mathrm{ax}$ is the scalar defined by formula
(\ref{proof of lemma final stage equation 2}).
It remains to observe that formulae
(\ref{proof of lemma final stage equation 10}),
(\ref{proof of lemma final stage equation 11})
imply (\ref{proof of lemma final stage equation 6}).
\end{proof}

The following theorem is the main result of our paper.

\begin{thm}
\label{main theorem}
In the special case with no dependence on $x^3$ a coframe
$\vartheta$ and a density $\rho$ are a solution of the field
equations for the Lagrangian density $L(\vartheta,\rho)$ if and only
if the corresponding spinor field is a solution of the field
equation for the Lagrangian density $L_\mathrm{Dir}^+(\xi)$ or the
field equation for the Lagrangian density~$L_\mathrm{Dir}^-(\xi)$.
\end{thm}

\begin{proof}
Denote by $L(\xi)$ the Lagrangian density (\ref{Our model equation 6})
but with $\vartheta$ and $\rho$ expressed via $\xi$. Accordingly, we
rewrite the factorisation formula (\ref{factorisation}) as
\begin{equation}
\label{factorisation modified}
L(\xi)=-\frac{32m}9
\frac{L_\mathrm{Dir}^+(\xi)L_\mathrm{Dir}^-(\xi)}
{L_\mathrm{Dir}^+(\xi)-L_\mathrm{Dir}^-(\xi)}\,.
\end{equation}
Observe also that the Dirac Lagrangian densities
$L_\mathrm{Dir}^\pm$ defined by formula~(\ref{Dirac Lagrangian plus minus})
possess the property of scaling covariance:
\begin{equation}
\label{proof of theorem equation 1}
L_\mathrm{Dir}^\pm(e^h\xi)=e^{2h}L_\mathrm{Dir}^\pm(\xi)
\end{equation}
where $h:M\to\mathbb{R}$ is an arbitrary scalar function.

We claim that the statement of the theorem
follows from (\ref{factorisation modified})
and
(\ref{proof of theorem equation 1}).
The proof presented below is an abstract one and does not depend
on the physical nature of the
dynamical variable $\xi$, the only requirement being that it is an element
of a vector space so that scaling makes sense.

Note that formulae
(\ref{factorisation modified})
and
(\ref{proof of theorem equation 1})
imply that the Lagrangian density $L$
possesses the property of scaling covariance,
so all three of our Lagrangian densities,
$L$, $L_\mathrm{Dir}^+$ and $L_\mathrm{Dir}^-$,
have this property.
Note also that if $\xi$ is a solution of the field equation for
some Lagrangian density $\mathcal{L}\,$
possessing the property of scaling covariance
then $\mathcal{L}(\xi)=0$. Indeed, let us perform a scaling
variation of our dynamical variable
\begin{equation}
\label{scaling variation}
\xi\mapsto\xi+h\xi
\end{equation}
where $h:M\to\mathbb{R}$ is an arbitrary ``small'' scalar function
with compact support. Then
$0=\delta\int\mathcal{L}(\xi)=2\int h\mathcal{L}(\xi)$
which holds for arbitrary $h$ only if $\mathcal{L}(\xi)=0$.

In the remainder of the proof the variations of $\xi$ are arbitrary
and not necessarily of the scaling type (\ref{scaling variation}).

Suppose that $\xi$ is a solution of the field equation for
the Lagrangian density $L_\mathrm{Dir}^+$.
[The case when $\xi$ is a solution of the field equation for
the Lagrangian density $L_\mathrm{Dir}^-$ is handled similarly.]
Then $L_\mathrm{Dir}^+(\xi)=0$ and, in view of (\ref{density is nonzero}),
$L_\mathrm{Dir}^-(\xi)\ne0$.
Varying $\xi$, we get
\begin{multline*}
\!\!\!\!
\delta\!\int\!\!L(\xi)
=-\frac{32m}9\Bigl(
\int\!
\frac{L_\mathrm{Dir}^-(\xi)}
{L_\mathrm{Dir}^+(\xi)\!-\!L_\mathrm{Dir}^-(\xi)}
\delta L_\mathrm{Dir}^+(\xi)
+\!
\int\!\!
L_\mathrm{Dir}^+(\xi)
\delta\frac{L_\mathrm{Dir}^-(\xi)}
{L_\mathrm{Dir}^+(\xi)\!-\!L_\mathrm{Dir}^-(\xi)}
\Bigr)
\\
=\frac{32m}9\int\delta L_\mathrm{Dir}^+(\xi)
=\frac{32m}9\,\delta\int L_\mathrm{Dir}^+(\xi)
\end{multline*}
so
\begin{equation}
\label{formula for variation of our action}
\delta\int L(\xi)=\frac{32m}9\,\delta\int L_\mathrm{Dir}^+(\xi)\,.
\end{equation}
We assumed that $\xi$ is a solution of the field equation for
the Lagrangian density $L_\mathrm{Dir}^+$ so
$\delta\int L_\mathrm{Dir}^+(\xi)=0$ and formula
(\ref{formula for variation of our action}) implies that
$\delta\int L(\xi)=0$. As the latter is true for an
arbitrary variation of $\xi$ this means that
$\xi$ is a solution of the field equation for the Lagrangian
density $L$.

Suppose that $\xi$ is a solution of the field equation for
the Lagrangian density $L$.
Then $L(\xi)=0$ and formula(\ref{factorisation modified})
implies that either $L_\mathrm{Dir}^+(\xi)=0$ or $L_\mathrm{Dir}^-(\xi)=0$;
note that in view of (\ref{density is nonzero}) we cannot have simultaneously
$L_\mathrm{Dir}^+(\xi)=0$ and $L_\mathrm{Dir}^-(\xi)=0$.
Assume for definiteness that $L_\mathrm{Dir}^+(\xi)=0$.
[The case when $L_\mathrm{Dir}^-(\xi)=0$ is handled similarly.]
Varying $\xi$ and repeating the argument from the previous paragraph
we arrive at (\ref{formula for variation of our action}).
We assumed that $\xi$ is a solution of the field equation for
the Lagrangian density $L$ so
$\delta\int L(\xi)=0$ and formula
(\ref{formula for variation of our action}) implies that
$\delta\int L_\mathrm{Dir}^+(\xi)=0$. As the latter is true for an
arbitrary variation of $\xi$ this means that
$\xi$ is a solution of the field equation for the Lagrangian
density $L_\mathrm{Dir}^+$.
\end{proof}

The proof of Theorem \ref{main theorem} presented above may appear
to be non-rigorous but it can be easily recast in terms of
explicitly written field equations.

\section{Discussion}
\label{Discussion}

The mathematical model formulated in Section \ref{Our model} is
based on the idea that every point of spacetime can rotate and that
rotations of different points are totally independent. The idea of
studying such continua belongs to the Cosserat brothers~\cite{Co}.
Recall that in classical elasticity the deformation of a continuum
is described by a (co)vector function $u$, the field of
displacements, which is the dynamical variable (unknown quantity) in
the system of equations. Displacements, of course, generate
rotations: the infinitesimal rotation caused by a displacement field
$u$ is $du$, the exterior derivative of $u$. The Cosserat brothers'
idea was to make rotations totally independent of displacements, so
that the coframe (field of orthonormal bases attached to points of
the continuum) becomes an additional dynamical variable.

Our model is a special case of Cosserat elasticity in that we model
spacetime as a continuum which cannot experience displacements, only
rotations. The idea of studying such continua is also not new: it
lies at the heart of the theory of \emph{teleparallelism}
(=~absolute parallelism), a subject promoted in the end of the 1920s by A. Einstein
and \'E. Cartan \cite{letters,unzicker-2005-,sauer}. It is interesting that
Einstein pursued this activity precisely with the aim of modelling the electron,
but, unfortunately, without success.

The differences between our mathematical model
formulated in Section \ref{Our model} and
mathematical models commonly used in teleparallelism are as follows.
\begin{itemize}
\item
We assume the metric to be prescribed (fixed) whereas in
teleparallelism it is traditional to view the metric as a dynamical
variable. In other words, in works on teleparallelism it is
customary to view (\ref{constraint for coframe}) not as a constraint
but as a definition of the metric and, consequently, to vary the
coframe without any constraints at all. This is not surprising as
most, if not all, authors who contributed to teleparallelism came to
the subject from General Relativity.
\item
We choose a very particular Lagrangian density (\ref{Our model equation 6})
containing only one irreducible piece of torsion (axial) whereas in
teleparallelism it is traditional to choose a more general
Lagrangian containing all three pieces
(tensor, trace and axial), see formula (26)
in \cite{cartantorsionreview}.
\end{itemize}

We now explain the motivation behind our choice of the
Lagrangian density~(\ref{Our model equation 6}).
Suppose for simplicity that we
don't have electromagnetism, i.e. that $A=0$, in which case the
extended coframe and extended metric agree (\ref{proof of lemma equation 2}).
Let us perform a conformal rescaling of the extended coframe:
${\bm{\vartheta}}{}^j\mapsto e^h{\bm{\vartheta}}{}^j$, $j=0,1,2,3,4$,
where $h:M\to\mathbb{R}$ is an arbitrary scalar function.
Then the metric and axial torsion scale as
$\mathbf{g}\mapsto e^{2h}\mathbf{g}$ and
\begin{equation}
\label{rescaling axial torsion}
\mathbf{T}^\mathrm{ax}\mapsto e^{2h}\mathbf{T}^\mathrm{ax}
\end{equation}
respectively. Here the remarkable fact is that the derivatives of $h$
do not appear in formula (\ref{rescaling axial torsion})
which means that axial torsion is the irreducible piece of torsion
which is conformally covariant. It remains to note that if we scale the
density $\rho$ as $\rho\mapsto e^{2h}\rho$ then the
Lagrangian density~(\ref{Our model equation 6}) will not change.

Thus, the guiding principle in our choice of the
Lagrangian density~(\ref{Our model equation 6}) is conformal invariance.
This does not, however, mean that our mathematical model
formulated in Section \ref{Our model} is conformally invariant:
formula (\ref{Our model equation 2 point 5}) does not allow for
conformal rescalings. The Kaluza--Klein extension is a
procedure which breaks conformal invariance, as one would expect when
introducing mass.

The main result of our paper is Theorem \ref{main theorem}
which establishes that in the
special case with no dependence on $x^3$
our mathematical model is equivalent to the Dirac equation.
This special case is known in literature as the Dirac equation
in dimension $1+2$ and is in itself the subject of extensive research.

This leaves us with the question what can be said about the general
case, when there is dependence on all spacetime coordinates
$(x^0,x^1,x^2,x^3)$. In the general case our model is clearly not
equivalent to the Dirac equation because it lacks one real
dynamical degree of freedom,
see last paragraph in Section~\ref{Choosing a common language}.
Our plan for the future is to examine \emph{how much} our model differs
from the Dirac model in the general case. We plan to compare
the two models by calculating energy spectra of the electron in
a given stationary electromagnetic field, starting with the case of the Coulomb
potential (hydrogen atom).

The spectral-theoretic analysis of our model will, however, pose a
monumental analytic challenge. There are several fundamental
issues that have to be dealt with.
\begin{itemize}
\item
Our model does not appear to fit into the standard scheme of
strongly hyperbolic systems of partial differential equations.
\item
The eigenvalue (= bound state) problem for our model is nonlinear.
\item
Our construction relies on the density $\rho$ being strictly
positive. This assumption may fail when one seeks bound states other
than the ground state.
\end{itemize}

\end{document}